# Ionization Cooling for Muon Experiments*

Y. Alexahin, D. Neuffer, E. Prebys (FNAL APC)

## 1. Introduction

There are a number of experiments that could benefit from using cooled muon beams [1]:

- flavor violating μ→e conversion in muonic atoms ("mu2e"),
- other charged lepton flavor violation experiments: μ→eγ, μ→eee,
- muon electric dipole moment measurement,
- precision muon lifetime measurement,
- muonium to anti-muonium conversion

and many more, including experiments using neutrinos from muon decay. Besides experiments with muons and neutrinos, ionization cooling can also be used for beams of β-unstable elements (β-beams) [2].

## 2. Mu2e experiment requirements

Here we will discuss in some detail the possibilities offered by muon ionization cooling for the next generation of the mu2e experiment, which is the flagship of the FNAL muon program.

The goal is to achieve a single event sensitivity of a few times $10^{-19}$ and to be able to explore the use of different target materials, from Al, where the muonic atom lifetime is ~0.9 μs, to Au, where the lifetime is only 77ns. Therefore the muon beam source must satisfy the following requirements:

1. A flux of the order of $10^{12}$ muons/s.
2. There must be no muons with momentum > 76 MeV/c, since they could produce electrons at the conversion energy. Also the muons must be stopped in 400 microns or less of Al or the equivalent thickness of other material. This means that muons must have momentum < 20 MeV/c.
3. The width of the muon pulse should be much shorter than the muon lifetime in the target material, 10ns seems adequate for both Al and Au
4. The time between muon pulses within a train must be long enough compared with the pulse width but should not exceed ~3 muon lifetimes in the target material.
5. There must be no pions and antiprotons. These are naturally removed with the use of a long cooling channel.

Requirements 2 and 3 determine what the muon beam longitudinal emittance should be. Let us assume for the average momentum and the r.m.s. values of momentum and time spread $p_{fin}$=15MeV/c, $\sigma_{pfin}$=2MeV/c, $\sigma_{tfin}$=2ns so that muons at ±2.5σ lie within the specified ranges. Then the normalized longitudinal emittance is

$$\varepsilon_{\|N} = \sigma_{pfin}\sigma_{zfin} / m_\mu = \sigma_{pfin}\sigma_{tfin} v_z / m_\mu = 1.6\, mm \qquad (1)$$

Such emittance can be obtained with the 6D cooling channel designed for a muon collider [3]. Actually a beam with much higher emittance can be used if the tails of the distribution are cut off at the expense of the intensity.

___________________________________________
*) Work supported by Fermi Research Alliance, LLC under Contract DE-AC02-07CH11359 with the U.S. DOE.

There is no particular requirement on the transverse emittance so far, but if we want to keep the beam size $\sigma_\perp<5$cm in a 1T solenoidal field ($\beta_\perp=0.1$m) then

$$\varepsilon_{\perp N} = \sigma_\perp^2 / \beta_\perp \times p_{fin} / m_\mu \leq 3.5\ mm, \tag{2}$$

which is somewhat larger than the transverse emittance expected from the Initial 6D Cooling Channel (see Table 1).

## 3. Proton Driver

The Fermilab Booster is being upgraded to work at a 15Hz repetition rate. With $6\cdot10^{12}$ of protons per cycle (this number has been achieved in studies, but more work is required to reduce the emittance growth) there will be $0.9\cdot10^{14}$ pps. With completion of PIP-II, the number of protons per cycle will increase to $7\cdot10^{12}$ while the repetition rate will be increased to 20Hz, providing a total flux $1.4\cdot10^{14}$ pps. However, we will use the former number for estimates.

One Booster cycle gives a batch of 80 bunches separated by 19ns, which is too short a time even for experiments with the Au target. Therefore the batch should be transferred to a "waiting ring" from where the bunches will be disbursed either one-by-one or – if the detector is able to handle a large number of tracks – in groups of a few bunches to be combined on the muon production target.

The Debuncher where the revolution period is 1.69μs can be used as such a storage ring. With one-by-one bunch extraction at every other turn the distance between the bunches will be 3.4μs, the total length of the batch will be ~270μs and, with a 15Hz repetition rate, the duty factor of the subsequent channel will be 0.4%.

## 4. Ionization Cooling

Since the cooling channel will have to work with a relatively high duty factor (or even in CW mode if the Project-X linac will be used as the driver), it is important to use an open-cell RF cavity design, since losses in Be windows would be prohibitively high. In order not to lose much in RF gradient we could fill the cavities with high pressure hydrogen gas. This option is assumed below. However, vacuum RF can also be used [4].

Let us take for example the front end and the HPRF cooling channel developed for the muon collider [3]. Table 1 presents the basic parameters of the front end, the initial cooling section (HFOFO snake) and the first two stages of the Helical Cooling Channel (HCC): length, normalized emittances, transmission factors and the expected total number of μ⁻ in the best 20 consecutive bunches per 8GeV proton.

Table 1. Parameters of the MC cooling channel [3]

| step | L (m) | $\varepsilon_{\perp N}$ (mm) | $\varepsilon_{\parallel N}$ (mm) | T (%) | $N_{\mu-}/N_p$ |
|---|---|---|---|---|---|
| Front End | 102 | 16 | 40 | - | 0.16 |
| 325 MHz HFOFO snake | 130 | 2.7 | 7.2 | 70 | 0.11 |
| Matching into HCC | 12 | 3.3 | 7.7 | 72 | 0.08 |
| 325 MHz HCC | 70 | 1.6 | 2.4 | 90 | 0.07 |
| 650 MHz HCC | 120 | 0.8 | 1.2 | 80 | 0.06 |

The Helical Cooling Channel provides significantly smaller emittances than those required by experiments but at the price of reduced transmission and much higher complexity and cost of the channel. But as long as a very low transverse emittance is not needed we can use only the initial cooling section (HFOFO snake). Then to obtain the required muon bunch length and momentum spread the tails of the longitudinal distribution must be cut off using collimators in high dispersion locations and/or fast kickers. The overall hit on the intensity can be as high as 60% leaving ~ 0.045 $\mu^-$ per one 8GeV proton in 20 bunches. With the proton driver flux of $0.9 \cdot 10^{14}$ pps we get $4 \cdot 10^{12}$ $\mu^-$/s.

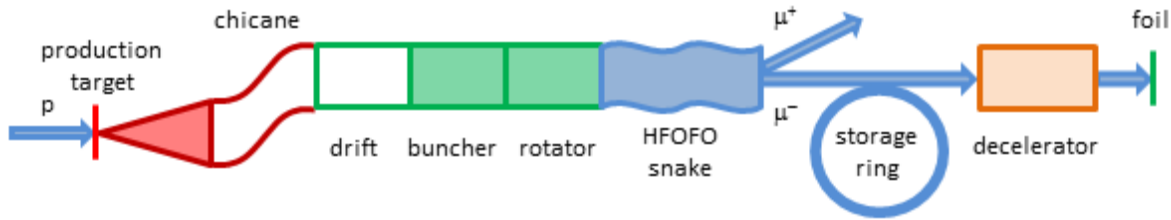

Figure 1: Schematic of the high-intensity muon source

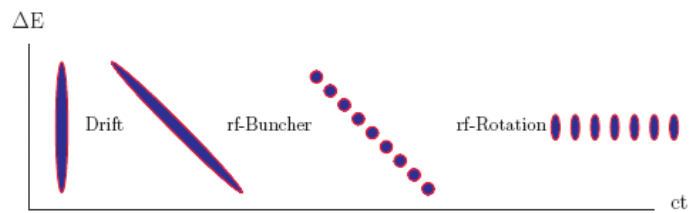

Figure 2: Front End concept

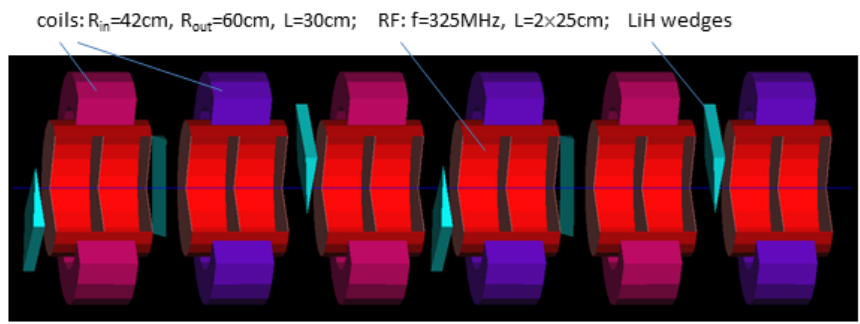

Figure 3: Schematic of one period of the Helical FOFO snake. The solenoid polarity is indicated by the alternating colors. The inclination of the solenoids is too small to be visible.

## 5. Muon Source Concept

The proposed scheme of the high-intensity muon source is shown in Fig.1. Proton beam from the driver (as discussed in Section 3) hits a target immersed in a strong magnetic field (15-20T) and produces pions which decay into muons in the downstream region. The expanding magnetic field transforms their transverse momentum into longitudinal. The chicane directs muons and pions with useful momentum

(below 1GeV/c) into the drift section where the time-energy correlation is established. Higher energy particles (including protons) proceed to a beam dump.

In the Front End muons of both signs are bunched as illustrated by Fig.2 and then cooled in the HFOFO snake (Fig.3). With the current version of the Front End developed for the MC cooling channel [5], the muons produced by a 2-3ns short proton bunch are distributed among more than 30 bunches of each sign so that the length of the whole train exceeds 100ns.

In Table 1 we cite the total number of muons in the 20 best bunches assuming they all can be used. The major difficulty with this arises from bunch lengthening during deceleration. To avoid bunch overlap the cooled muons can be placed in a small storage ring and dispatched one-by-one with time interval ~170ns to the deceleration channel and the conversion target. There will be 1600 muon bunches generated by one Booster batch following each other with frequency 5.9MHz.

The last part of the muon source is the deceleration channel. With the present HFOFO snake design [5] the average muon momentum is lowered from 250MeV/c to 200MeV/c along the channel. Probably the deceleration with simultaneous ionization cooling can go on further but not by much since the longitudinal anti-damping due to the dependence of ionization losses on momentum becomes prohibitive at <~150MeV/c.

The remaining deceleration must be carried out by an RF field. To provide the required small momentum spread the RF pulse profile should be quite linear over the bunch full length which increases in the course of deceleration to 10ns. High repetition rate of 5.9MHz necessitates the use of multi-frequency RF excluding application of induction columns.

Assuming (optimistically) the average decelerating gradient of 5MV/m the muon losses can be estimated at 5% over ~40m. With account for these losses and decays in the storage ring the muon flux on the muon capture target will be $2.6 \cdot 10^{12}$ $\mu^-$/s which is enough to reach single event sensitivity of $5 \cdot 10^{-19}$ in 7 months for Al, assuming the product of acceptance and reconstruction efficiency is 10%. Thus, the cooling channel approach shows the possibility of developing greatly increased intensity useful for a second generation of experiments. This concept can be further adapted and improved in future designs that are more closely aligned to the final goals and capabilities of mu2e experiments .